\documentclass{appolb}
\usepackage{graphicx}
\usepackage{amsfonts,amssymb,amsmath,bm}
\usepackage{cite}

\begin{document}
\title{Gluon mass through massless bound-state excitations.%
\thanks{Presented at the Workshop ``Excited QCD 2012''. Peniche, Portugal, May 6-12,2012.}%
}
\author{David Ibanez
\address{Department of Theoretical Physics and IFIC,\\ 
        University of Valencia-CSIC,\\
        E-46100, Valencia, Spain.\\
        E-mail: David.Gil@ific.uv.es}
\\
}
\maketitle
\begin{abstract}
Recent large-volume lattice simulations have established that, in the Landau gauge, the gluon propagator is infrared-finite. 
The most natural way to explain this observed finiteness is the generation of a nonperturbative, momentum-dependent gluon mass. 
Such a mass may be generated gauge-invariantly 
by employing the Schwinger mechanism, whose main assumption is the  dynamical formation of  
massless bound-state excitations. In this work we demonstrate that this key assumption is indeed realized 
by the QCD dynamics. Specifically, the Bethe-Salpeter equation describing the aforementioned massless excitations 
is derived and solved under certain approximations, and non-trivial solutions are obtained.
\end{abstract}
  
\section{Introduction}

It is by now a well-established fact that 
large-volume lattice simulations in the Landau gauge yield a 
gluon propagator that reaches a finite non-vanishing value 
in the deep 
infrared~\cite{Cucchieri:2009zt,Bogolubsky:2007ud,Bowman:2007du,Oliveira:2009eh}. 
Without a doubt, the most physical way of explaining this observed 
finiteness is to invoke the mechanism of dynamical gluon mass generation, first 
introduced in the seminal work of Cornwall~\cite{Cornwall:1981zr}, 
and subsequently studied in a series of articles~\cite{Aguilar:2006gr,Aguilar:2008xm,arXiv:1107.3968}. 
In this picture  
the fundamental Lagrangian of the Yang-Mills theory (or that of QCD) remains unaltered, and   
the generation of the gluon mass takes place dynamically, 
through the well-known Schwinger 
mechanism~\cite{Schwinger:1962tn,Jackiw:1973tr,Eichten:1974et},  
without violating any of the underlying symmetries (for related contributions see, e.g., \cite{Oliveira:2010xc}).

The main purpose of this presentation is to report on recent work~\cite{Aguilar:2011xe}, where the Schwinger mechanism in quarkless QCD has been examined. Specifically,  
the entire mechanism of gluon mass generation hinges on the 
appearance of massless poles inside the nonperturbative three-gluon vertex, 
which enters in the one-loop dressed diagram Fig.~\ref{gSDE} of the Schwinger-Dyson equation (SDE) governing the gluon propagator.
These poles correspond to the propagator of the scalar massless excitation, 
and interact with a pair of gluons through a very characteristic proper vertex, 
which, of course, must be non vanishing, or else the entire construction is invalidated.
The way to establish the existence of this latter vertex is by finding non-trivial solutions
to the homogeneous Bethe-Salpeter equation (BSE) that it satisfies.

\section{Gluon mass and the BS wave-function.}
\begin{figure}[!t]
\begin{center}
\includegraphics[scale=0.6]{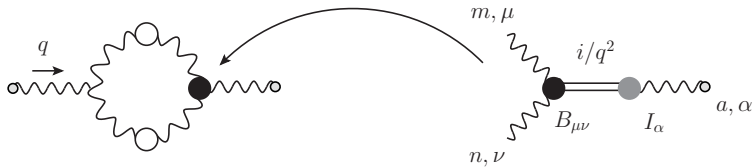}
\caption{\label{gSDE} Pictorial representation of the Schwinger mechanism in SDE. White (black) circles denote fully dressed propagators (vertices); a gray circle attached to the external legs indicates that they are background gluons.}
\end{center}
\end{figure}

In the formalism provided by the synthesis of the pinch technique (PT) with the background field method (BFM), known in the literature as the PT-BFM scheme~\cite{Binosi:2009qm}, the Schwinger mechanism is integrated into the SDE of the gluon propagator, given in the Landau gauge by\footnote{The usual transverse projector is defined as $P_{\mu\nu}(q) = g_{\mu\nu} - q_\mu q_\nu/q^2$.}
\begin{equation}\label{propagator}
\Delta_{\mu\nu}^{ab}(q) = \delta^{ab}\Delta_{\mu\nu}(q)\quad ; \quad \Delta_{\mu\nu}(q) = -iP_{\mu\nu}(q)\Delta(q^2),
\end{equation}
through the form of the three gluon vertex $\widetilde\Gamma_{\alpha\mu\nu}$, connecting one background with two quantum gluons. This is accomplished by adding a new non perturbative piece to the three gluon vertex, which can be cast in the form of Fig.~\ref{gSDE}, by setting
\begin{equation}\label{Uvertex}
\widetilde{U}_{\alpha\mu\nu}^{amn}(q,r,p) = \widetilde{I}_\alpha^{ab}(q) \bigg(\frac{i}{q^2}\delta^{bc}\bigg)B_{\mu\nu}^{cmn}(q,r,p).
\end{equation}
In this expression, the non perturbative quantity
\begin{equation}\label{Bexpansion}
B_{\mu\nu}(q,r,p) = B_1 g_{\mu\nu} + B_2 q_\mu q_\nu + B_3 p_\mu p_\nu + B_4 r_\mu q_\nu + B_5 r_\mu p_\nu,
\end{equation}
is the effective vertex describing the interaction between the massless excitation and two gluons. $B_{\mu\nu}$ is to be identified with the ``bound-state wave function'' (or ``BS wave function'') of the two-gluon bound-state, which, as we will see shortly, satisfies a homogeneous BSE. In addition, $i/q^2$ is the propagator of the scalar massless excitation. Finally, $\widetilde{I}_\alpha(q)$ is the (nonperturbative) transition amplitude, allowing the mixing between a (background) gluon and the massless excitation.

One can show~\cite{Aguilar:2011xe} that, in the limit of zero momentum transfer $q^2=0$, the relevant quantity to consider is the form factor $B_1$ appearing in Eq. (\ref{Bexpansion}). Furthermore, due to Bose symmetry with respect to the interchange $\mu\leftrightarrow\nu$ and $p\leftrightarrow r$, we must have $B_1(q,r,p) = -B_1(q,p,r)$, which implies in the aforementioned kinematical limit that $B_1(0,-p,p)=0$. So, when $q\rightarrow 0$, what survives is the derivative $B_1'$, which can be related to the derivative of the effective gluon mass through the exact all-order relation,
\begin{equation}\label{massrelation}
[m^2(p)]' = - \widetilde{I}(0) B_1'(p).
\end{equation}

\section{\label{bse}The Bethe-Salpeter equation}

\begin{figure}[!t]
\center{\includegraphics[scale=0.35]{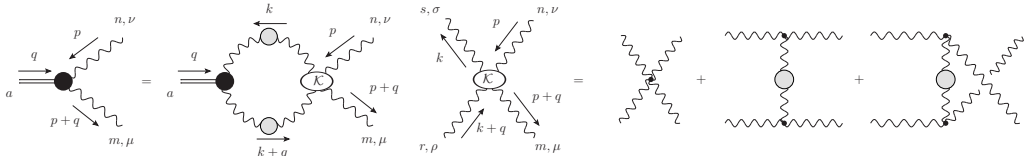}}
\caption{The BSE for the bound-state wave function $B_{\mu\nu}$ and the Feynman diagrams considered for the BS kernel.}
\label{RegPole}
\end{figure}

The existence of $B'_1$ is of paramount 
importance for the mass generation mechanism; essentially, the question 
boils down to whether or not   
the dynamical formation of a massless 
bound-state excitation of the type postulated above is possible.
As is well-known, in order to establish the 
existence of such a bound state one must 
{\bf (i)} derive the appropriate BSE for the 
corresponding bound-state wave function, $B_{\mu\nu}$,  
(or, in this case, its derivative),   
and {\bf (ii)} find non-trivial solutions for this integral equation.
An approximate BSE for the bound-state wave function $B_{\mu\nu}$ is given by the following expression [see Fig.~\ref{RegPole}]
\begin{equation}
B_{\mu\nu}^{amn} = 
\int_k B_{\alpha\beta}^{abc}\Delta^{\alpha\rho}_{br}(k+q)\Delta^{\beta\sigma}_{cs}(k){\cal K}_{\sigma\nu\mu\rho}^{snmr} \,.
\label{BS}
\end{equation}
As explained in~\cite{Aguilar:2011xe}, Eq. (\ref{BS}) is obtained from the full BSE satisfied by the vertex $\widetilde\Gamma_{\alpha\mu\nu}$.
We will next approximate the  four-gluon BS kernel ${\cal K}$ by the 
lowest-order set of diagrams shown in Fig.~\ref{RegPole}, where the vertices are bare, while 
the internal gluon propagators are fully dressed.
Going to Euclidean space, we define $x\equiv p^2$, $y\equiv k^2$, and $z\equiv (p+k)^2$; 
then, following the procedure outlined in~\cite{Aguilar:2011xe},
the BSE becomes
\begin{eqnarray}\label{34}
B'_1(x) &=& -\frac{\alpha_s C_A}{12\pi^2}\int_0^\infty dy yB'_1(y)\Delta^2(y)\sqrt{\frac{y}{x}} \nonumber \\
&\times & \int_0^\pi d\theta\sin^4\theta\cos\theta\bigg[z+10(x+y)+\frac{1}{z}(x^2+y^2+10xy)\bigg]\Delta(z)\,.
\label{euclideanBS}
\end{eqnarray}

As a further simplification, we approximate the gluon propagator $\Delta(z)$ 
appearing in the BSE of (\ref{euclideanBS}) [but not the $\Delta^2(y)$]
by its tree level value, that is, $\Delta(z)=1/z$. Then, the angular integration may be carried out 
exactly, yielding  
\begin{eqnarray}
B'_1(x) &=& \frac{\alpha_s C_A}{24\pi}\bigg\lbrace\int_0^x dyB'_1(y)\Delta^2(y)\frac{y^2}{x}
\bigg(3+\frac{25}{4}\frac{y}{x}-\frac{3}{4}\frac{y^2}{x^2}\bigg) \nonumber \\
&+& \int_x^\infty dyB'_1(y)\Delta^2(y)y\bigg(3+\frac{25}{4}\frac{x}{y}-\frac{3}{4}\frac{x^2}{y^2}\bigg)\bigg\rbrace \,.
\label{weakBS}
\end{eqnarray}

\section{Numerical analysis}
\begin{figure}[!t]
\begin{minipage}[b]{0.45\linewidth}
\noindent
\centering
\hspace{-1cm}
\includegraphics[scale=0.45]{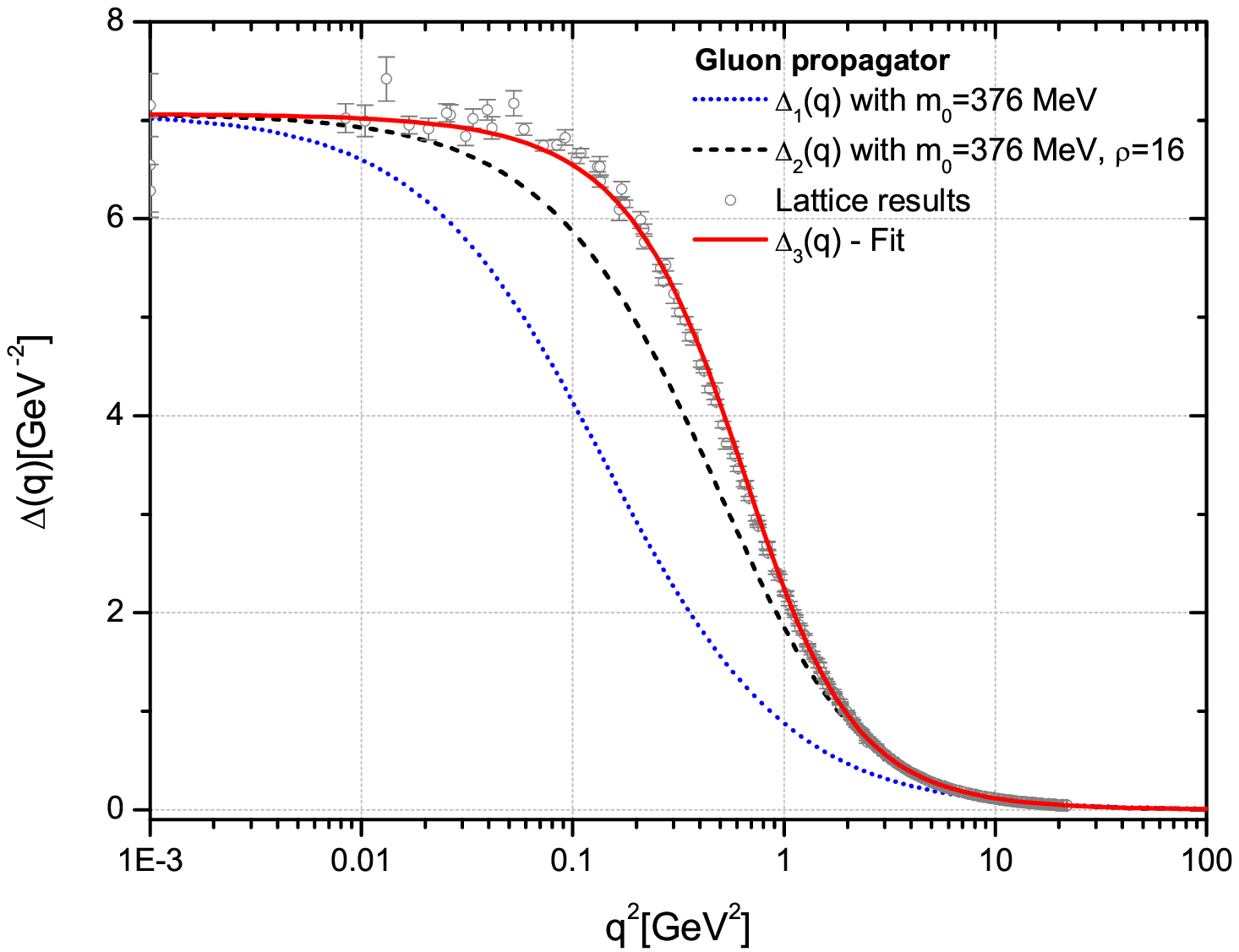}
\end{minipage}
\begin{minipage}[b]{0.45\linewidth}
\hspace{-0.5cm}
\noindent
\includegraphics[scale=0.45]{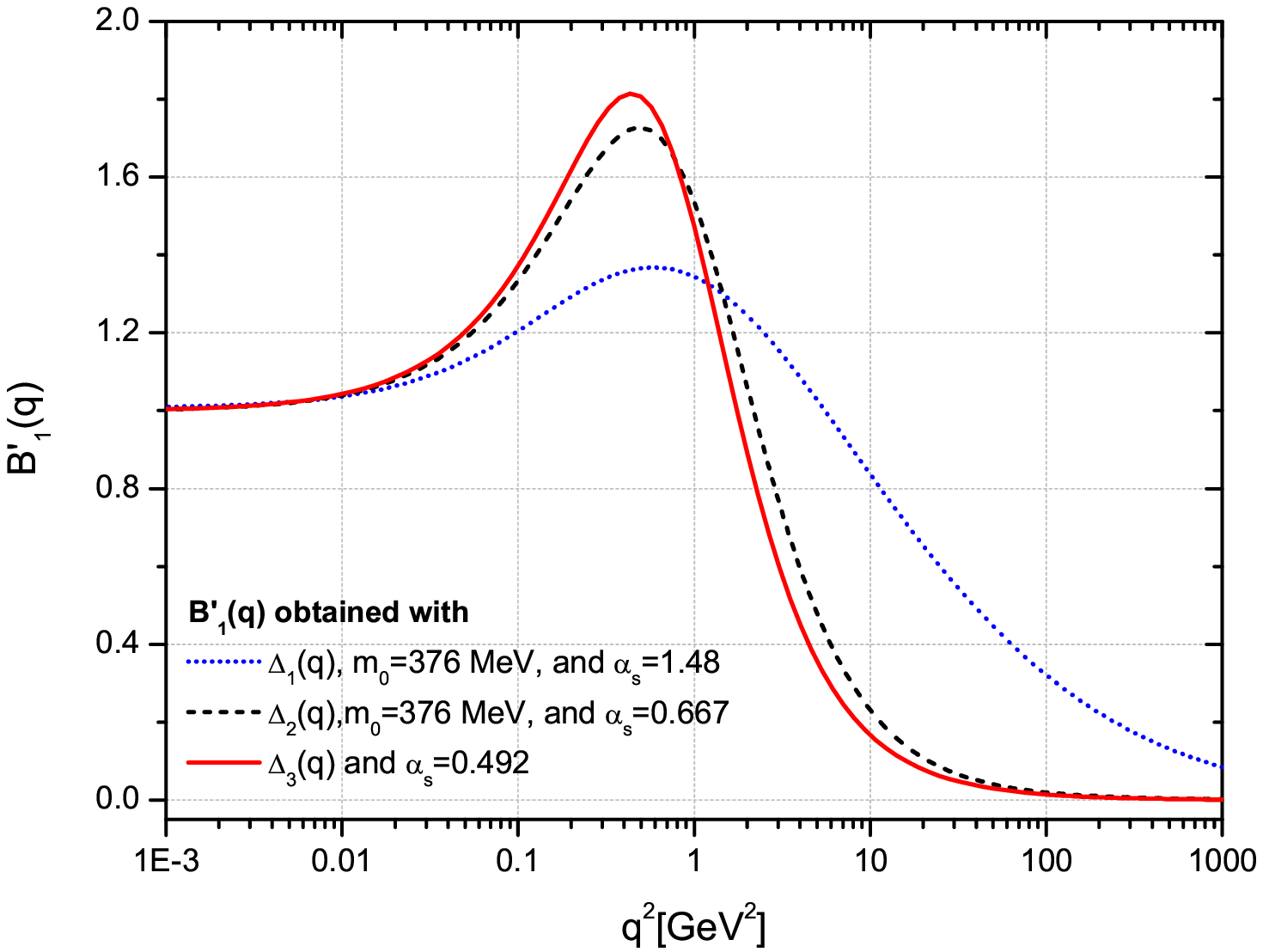}
\end{minipage}
\caption{The three models for the gluon propagator (left) and 
the corresponding solutions of the BS equation for $B'_1(x)$ (right).}
\label{props}
\end{figure}

Next we discuss the numerical solutions for 
Eq.~(\ref{weakBS}) for arbitrary values of $x$. 
Evidently, the main ingredient entering into its kernel is
the nonperturbative gluon propagator, $\Delta(q)$. In order to explore the sensitivity of 
the solutions on the details of $\Delta(q)$, 
we  will employ three infrared-finite forms focusing on their differences in the intermediate and asymptotic 
regions of momenta, given by
\begin{eqnarray}\label{smassive}
\Delta_1^{-1}(q^2) &=& q^2 + m^2_0 \,, \\
\Delta_2^{-1}(q^2) &=&  q^2\left[1+ \frac{13C_{\rm A}g^2}{96\pi^2} \ln\left(\frac{q^2 +\rho\,m_0^2}{\mu^2}\right)\right] +  m^2_0 \,,\label{gluon2} \\
\Delta_3^{-1}(q^2) &=& m_g^2(q^2) + q^2\left[1+ \frac{13C_{\rm A}g_1^2}{96\pi^2} \ln\left(\frac{q^2 +\rho_1\,m_g^2(q^2)}{\mu^2}\right)\right]\label{gluon3}.
\end{eqnarray}
The first expression (\ref{smassive}) corresponds to a tree-level massive propagator, where $m^2_0$ is a hard mass treated as a free parameter. On the left panel of Fig.~\ref{props}, the (blue) dotted curve represents $\Delta_1$ for $m_0=376 \,\mbox{MeV}$. The second model (\ref{gluon2}) contains the renormalization-group logarithm next to the momentum $q^2$, where $\rho$ is an adjustable parameter varying in the range of $\rho \in [2,16]$. Notice that the hard mass $m_0^2$
appearing in the argument of the perturbative logarithm acts as an infrared cutoff. The (black) dashed line represents Eq.~(\ref{gluon2}) when $\rho=16$, $m_0=376 \,\mbox{MeV}$, and $\mu=4.3$ GeV. Finally, Eq. (\ref{gluon3}) is simply a physically motivated fit for the gluon propagator determined by the large-volume lattice simulations of Ref.~\cite{Bogolubsky:2007ud}, where $m^2_g(q^2) = m^4/(q^2 + \rho_2 m^2)$ is a a running mass. The values of the fitting parameters are  
\mbox{$m= 520$\,\mbox{MeV}}, \mbox{$g_1^2=5.68$}, \mbox{$\rho_1=8.55$} and, \mbox{$\rho_2=1.91$}, and the (red) continuous line on the left panel of Fig.~\ref{props} represents this propagator. Notice that, in all three cases, we have fixed the value of $\Delta^{-1}(0)=m_0^2\approx 0.14$.

Our main findings may be summarized as follows.

{\it (a)} In Fig.~\ref{props}, right panel, we show the solutions 
of Eq.~(\ref{weakBS}) obtained using as input 
the three propagators shown on the left panel.
For the simple massive propagator
of Eq.~(\ref{smassive}), a solution for $B'_1$ is found for \mbox{$\alpha_s=1.48$};
in the case of $\Delta_2$ given by  Eq.~(\ref{gluon2}), a solution 
is obtained when \mbox{$\alpha_s=0.667$}, while for the lattice
propagator $\Delta_3$ of  Eq.~(\ref{gluon3}) a non-trivial 
solution is found when \mbox{$\alpha_s=0.492$}.

{\it (b)} Note that, due to the fact that Eq.~(\ref{weakBS}) is homogeneous and (effectively) linear,
if $B'_1$ is a solution then the function $cB'_1$ is also a solution, 
for any real constant $c$. Therefore,
the solutions shown on the right panel of Fig.~\ref{props} corresponds to  
a representative case of a family of  possible solutions, where the constant $c$ was chosen 
such that $B'_1(0)=1$. 

{\it (c)} Another interesting feature of the solutions of Eq.~(\ref{weakBS}) is  
the dependence of the observed peak on the support of the gluon propagator 
in the intermediate region of momenta. Specifically, 
an increase of the support of the gluon propagator in the approximate range (0.3-1) GeV 
results in a more pronounced peak in  $B'_1$.  

{\it (d)} In addition, observe that due to the presence of the 
perturbative logarithm 
in the expression for $\Delta_2$ and $\Delta_3$, the corresponding solutions $B'_1$  
fall off in the ultraviolet region much faster than those obtained using the 
simple $\Delta_1$ of Eq.~(\ref{smassive}). 

\section{Conclusions}

In this presentation we have 
reported recent progress~\cite{Aguilar:2011xe} on the study of the Schwinger
mechanism in QCD, which endows gluons with a dynamical mass in a gauge invariant way. 
This mechanism relies on the existence of massless bound-state excitations, 
whose dynamical formation is controlled by a homogeneous BSE.
As we have seen, under certain simplifying assumptions, this 
equation admits non-trivial solutions, thus furnishing additional 
support in favor of the specific mass generation mechanism described 
in a series of earlier works~\cite{Aguilar:2006gr,Aguilar:2008xm,arXiv:1107.3968}. 
In the future, this framework may serve as a starting point towards a consistent first-principle treatment of glueballs as bound states of gluons, at nonvanishing momenta. Specifically, one may envisage a combined BS and SDE analysis, in the spirit of similar approaches in the case of color singlets $q\bar{q}$ bound states, employed in the literature for the dynamical description of mesons.

{\it Acknowledgments:} 

This research is supported by the European FEDER and Spanish MICINN under grant FPA2008-02878.

\end{document}